\documentclass[letterpaper]{jpconf}
\usepackage{graphicx}
\usepackage{subfigure}

\def\pt{$P_T$}

\newcommand{\ppbar}{$p\bar{p}$}
\newcommand{\hbb}{$h \rightarrow b\bar{b}$}
\newcommand{\bbbar}{$b\bar{b}$}
\newcommand{\fb}{fb$^{-1}$}

\newcommand{\WHlvbb}{$Wh \rightarrow \ell \nu b\bar{b}$}
\newcommand{\ZHllbb}{$Zh \rightarrow \ell \ell  b \bar{b}$}
\newcommand{\ZHvvbb}{$Zh \rightarrow \nu \nu  b \bar{b}$}
\newcommand{\Mh}{$m_{h}$}
\newcommand{\gevcc}{GeV/$c^2$}

\newcommand{\MET}{\mbox{$\raisebox{.3ex}{$\not\!$}E_T$}}

\def\hgg{$h~\rightarrow \gamma \gamma~$}
\newcommand{\ptgg}{$P_{T}^{~\gamma \gamma}$}
\newcommand{\BRhgg}{$\cal{B}$r($h \rightarrow \gamma \gamma$)}
\newcommand{\HWW}{$h \rightarrow WW$}
\newcommand{\HWWlvlv}{$h \rightarrow WW \rightarrow \ell \nu \ell \nu$}

\begin{document}
\title{Higgs Boson Searches at CDF}

\author{Craig Group}

\address{Fermilab, Batavia, IL, USA\\
(On behalf of the CDF Collaboration)}

\begin{abstract}
Results are presented on searches for standard model and non-standard
model production of a Higgs boson in \ppbar\ collisions at
$\sqrt{s}= 1.96$ TeV with the CDF II detector at the Fermilab
Tevatron. Using data corresponding to 2-3.6 1/fb of integrated
luminosity, searches are performed in a number of different production
and decay modes.  No excess in data above that expected from
backgrounds is observed; therefore, we set upper limits on the
production cross section times branching fraction as a function of the
Higgs boson mass.

\end{abstract}

\section{Introduction}

    Although the Higgs mechanism~\cite{Higgs:1964pj} was proposed in the
   1960's, the fundamental particle it predicts, the Higgs boson ($h$), has
   yet to be discovered.  Direct limits from the LEP experiments
   exclude Higgs boson masses below
   $114.4$~\gevcc~\cite{Barate:2003sz} at 95\% Confidence Level (CL),
   while electroweak precision measurements place an indirect upper
   limit on the mass of a SM Higgs boson of
   $154$~\gevcc~\cite{Collaboration:2008ub} at 95\% CL.

    Here we will summarize the status of the search for the Higgs
    boson using the the CDF II detector~\cite{Acosta:2004yw} to
    analyze the \ppbar\ collision data from the Fermilab Tevatron.

\section{Summary of Standard Model Search Efforts}

  For Higgs boson masses below $135$~\gevcc, \bbbar\ is the main decay
  mode~\cite{Djouadi:1997yw}.  For higher masses the Higgs boson
  decays primarily into a pair of $W$ bosons.  Due to these different
  decay modes, it is natural to also split the discussion of the
  analysis effort at CDF into the low mass and high mass categories.     

  The search for a Higgs boson is quite challenging due to the large
  backgrounds and small signal expectation.  In order to
  increase signal-background discrimination, analyses make maximal use
  of the information in each event by employing multivariate
  techniques to collect the discriminating power of multiple input
  variables into a single more powerful output variable.  When using these
  techniques it is crucial to prove that all of the multivariate inputs
  and the discriminant outputs are described well by the background
  models.  These checks are performed by comparing the background
  model with the data in various control regions carefully defined to
  test the modeling of major background components.  Over the years
  CDF has built confidence in their detector modeling and has had
  success with the use of multivariate techniques in SM analyses such as
  the recent evidence and observation of single top quark
  production~\cite{Aaltonen:2009jj,Aaltonen:2008sy}.

  The Higgs boson search strategy is to perform dedicated analyses for
  each distinct final state with a significant production rate and
  then combine these results to make a statement about the sensitivity
  of the CDF experiment to the SM Higgs boson.

\subsection{Low mass Higgs boson searches}

    In the \hbb\ decay, each $b$ quark fragments into a jet of hadrons
  and the Higgs boson signal appears as a peak in the invariant mass
  distribution of these two jets.  The two-jet signature alone is not
  useful to reject the dijet QCD background which is produced at the
  Tevatron with a rate about ten orders of magnitude higher than the
  Higgs boson.  To handle this problem the low mass Higgs boson
  searches focus on production processes where the Higgs boson is
  produced in association with a $W$ or $Z$ boson ($Vh$).
  
    The requirements of a high \pt\ charged lepton candidate, missing
  transverse energy (\MET), and at least one $b$-tagged jet reduce the
  background in the \WHlvbb~\cite{wh,whPRD} search channel.  In the
  past CDF has used an analysis based on an artificial neural network
  (NN) which includes the dijet invariant mass, the total system \pt\
  and, the event \pt\ imbalance to improve the discrimination between
  the Higgs signal and background.  Recently, CDF has combined this
  analysis with a new analysis which incorporates matrix element (ME)
  calculations into a boosted decision tree.  This
  combination was done using another NN optimized based on genetic
  algorithms~\cite{Whiteson:2006ws}.  The combined result is the most
  sensitive low mass analysis obtaining observed (expected) 95\%
  C.L. limit of 5.6 (4.8) times the SM prediction of the production
  cross section for a Higgs boson mass of 115~\gevcc\ using 2.7~\fb\
  of integrated luminosity..
 
    The \ZHllbb ~\cite{Aaltonen:2008wj} has a smaller background but
  also a smaller signal expectation do to the reduced cross section of
  $Zh$ production.  The reach of this analysis has been improved by
  loosening the lepton identification requirements to increase signal
  acceptance.  In addition, this final state is fully constrained by
  the reconstructed objects and the \MET\ can be used to correct the
  jet energies to improve the dijet mass resolution.  CDF has a
  two-dimensional NN analysis and an analysis based on ME
  probabilities.  For a Higgs boson mass of 115~\gevcc\ the NN
  analysis obtains an observed (expected) 95\% C.L. limit of 7.7 (9.1)
  times the SM prediction of the production cross section using
  2.7~\fb\ of integrated luminosity.
   
   The third major low mass Higgs analysis is based on a \MET\
   requirement and identifying $b$ jets but does not allow any
   reconstructed leptons in the events~\cite{Aaltonen:2008mi}.  It is
   sensitive to \ZHvvbb\ but also \WHlvbb\ when the lepton escapes
   detection.  Without the charged lepton requirement the dominant
   background is QCD events where mismeasured jets fake the \MET\
   requirement.  A NN selection tool is derived which uses information
   about the angular correlations between the jets and the \MET\ as
   well as a ``missing \pt'' variable based on tracking information.
   By cutting on this NN the signal to background ratio is improved to
   the level of the lepton based analyses, making this channel
   competitive as one of the most sensitive low mass Higgs boson
   search channels.  The \MET+$bb$ analysis obtains observed
   (expected) 95\% C.L. limit of 6.9 (5.6) times the SM prediction of
   the production cross section for a Higgs boson mass of 115~\gevcc\
   using 2.1~\fb\ of integrated luminosity.

\subsection{High mass Higgs boson searches}

  At high mass the \HWWlvlv\ channel dominates the sensitivity to the
Higgs boson.  The leptonic decay mode of the $W$ bosons is chosen to
reduce background and improve signal purity.  The CDF
analysis~\cite{Aaltonen:2008ec} includes all significant production
modes (gluon fusion, $Vh$, and vector boson fusion (VBF)), and splits
the analysis up based on the number of jets observed in the final
state.  This is useful since the background and signal composition
change considerably depending on the number of jets.  The \HWW\
analysis is the most sensitive single analysis at CDF.

\subsection{Results}

\begin{figure*}[th]
\begin{center}
\includegraphics[width=0.48\textwidth]{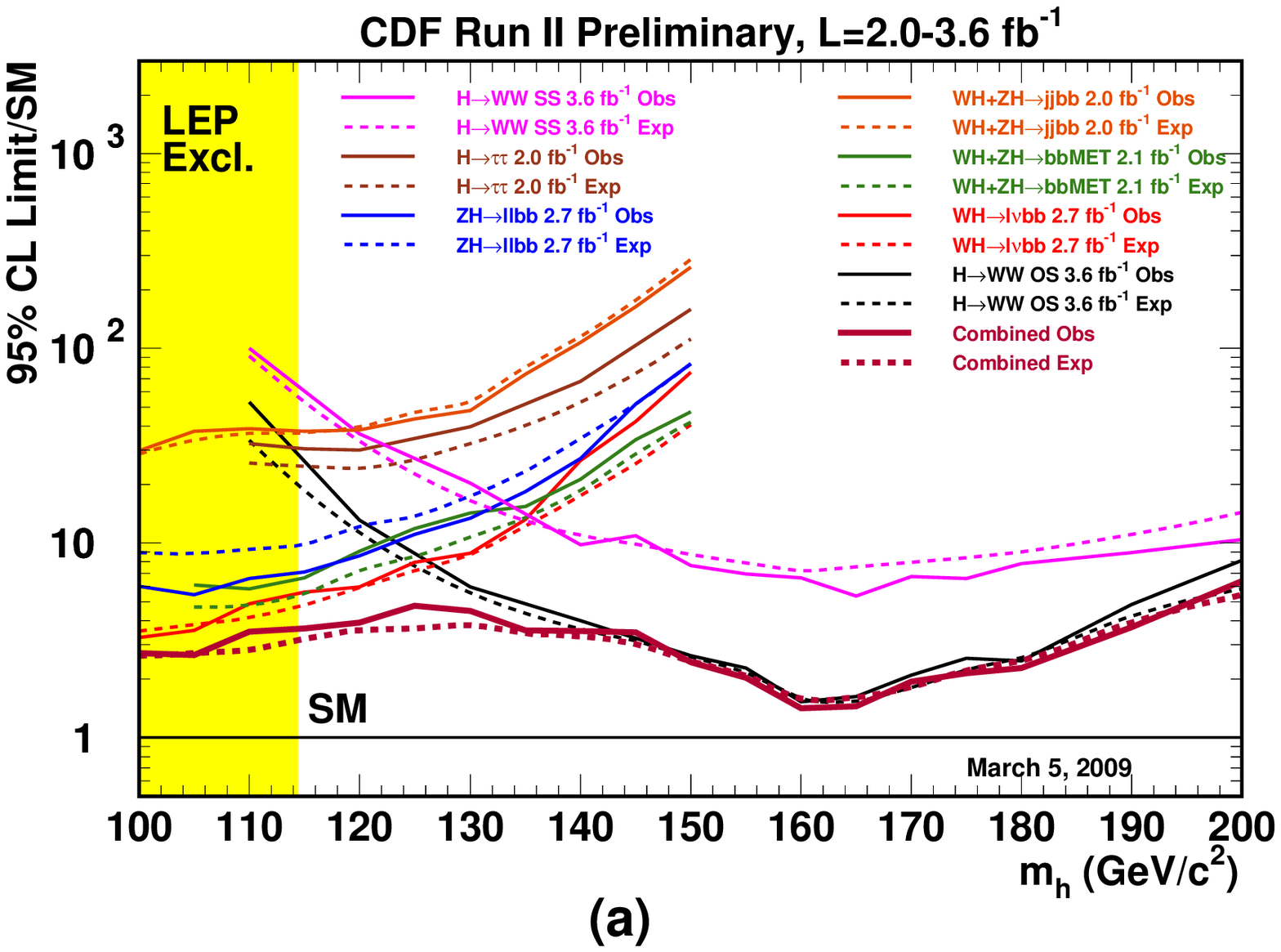}
\includegraphics[width=0.48\textwidth]{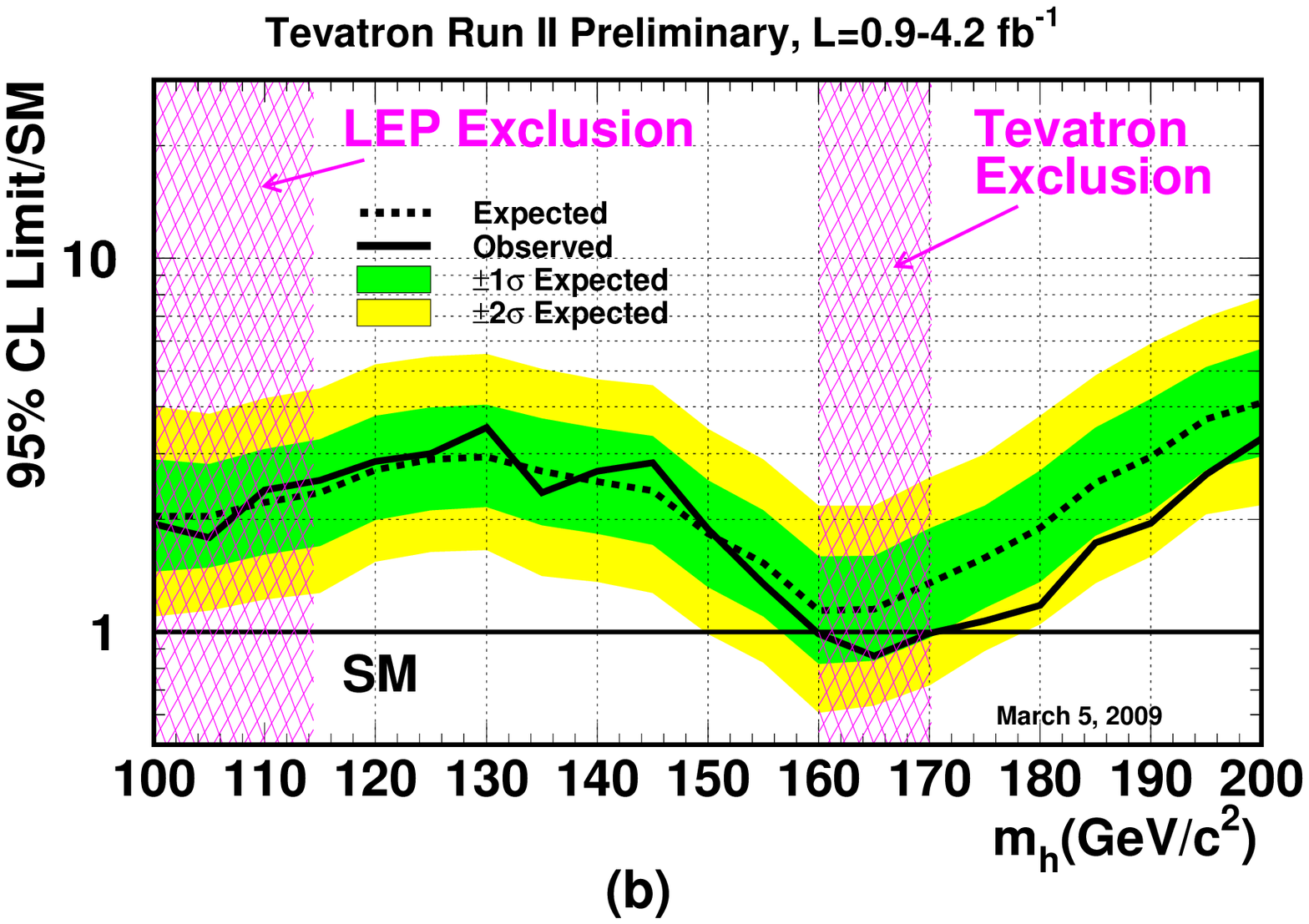}
%\begin{center}
%\subfigure[]{
%\includegraphics[width=0.48\textwidth]{figs/collectedlimits_mar5.eps}
%\label{fig:limitsa}
%}
%\subfigure[]{
%\includegraphics[width=0.48\textwidth]{figs/collectedlimits_CDF_D0.eps}
%\label{fig:limitsb}
%}
\caption[]{Limits at 95\% C.L. on the Higgs boson production cross
  section in factors away from the standard model prediction  for the individual analysis channels and the CDF combined
  result (a).  The full Tevatron result from the combination of the CDF
  and D\O\ experiment (b). }
\label{fig:CDF_limits}
\end{center}
\end{figure*}

  The observed (expected) limits on the Higgs boson cross section in
units of the SM prediction for all of the CDF analyses are shown in
Fig.~\ref{fig:CDF_limits}(a).  These results are combined into a single
limit on the Higgs boson production rate for each Higgs boson mass
hypothesis.  In addition to the most sensitive low mass analyses, CDF
also has analyses that focus on the all hadronic final state ($Vh$
where the vector boson decays into two jets), and an inclusive $h
\rightarrow \tau \tau$ analysis that are included in the combination.
The result of the combination of CDF results is shown as the garnet
line in Fig.~\ref{fig:CDF_limits}(a).  The limits range from 1.4 to
about 5 times the prediction of the SM rate.

   The CDF results have also been combined with the results of the
   D\O\ experiment.  This is shown in Fig.~\ref{fig:CDF_limits}(b).
   The combined result excludes Higgs bosons with masses between 160
   and 170 \gevcc\ at the 95\% C.L.  This is the first new exclusion
   of a standard model Higgs boson based on a direct search using
   Tevatron data.
   
\section{Beyond the Standard Model Example: Fermiophobic Higgs Boson Search}
The CDF experiment also searches for Higgs bosons produced in many
extensions to the SM.  One example which has been recently
updated~\cite{hgg} is a search in the diphoton final state.  The SM
prediction for the {$h \rightarrow \gamma \gamma$} branching fraction
is extremely small (reaching a maximal value of only about 0.2\% at a
Higgs boson mass (\Mh)$~\sim 120$~\gevcc)~\cite{Djouadi:1997yw};
however, in ``fermiophobic'' models, where the coupling of the Higgs
boson to fermions is suppressed, the diphoton decay can be greatly
enhanced.  This phenomenon has been shown to arise in a variety of
extensions to the
SM~\cite{Haber:1978jt,Gunion:1989ci,Basdevant:1992nb,Barger:1992ty,Akeroyd:1995hg},
and the resulting collider phenomenology has been
described~\cite{Dobrescu:1999gv,Matchev1,Mrenna}.  For this
fermiophobic case, the decay into the diphoton final state dominates
at low Higgs boson masses and is therefore the preferred search
channel.

A benchmark fermiophobic model is considered in which the Higgs boson
 does not couple to fermions, yet retains its SM couplings to bosons.
 In this model, the fermiophobic Higgs boson production is dominated
 by two processes: $Vh$, and VBF.  Two identified photon candidates
 are required in the analysis.  In addition, a cut is applied on the
 transverse momentum of the diphoton pair (\ptgg) designed to optimize
 sensitivity (VBF and $Vh$ have more \ptgg\ on average than the SM
 diphoton production and QCD backgrounds).

\begin{figure*}[th]
\begin{center}
\includegraphics[width=0.40\textwidth]{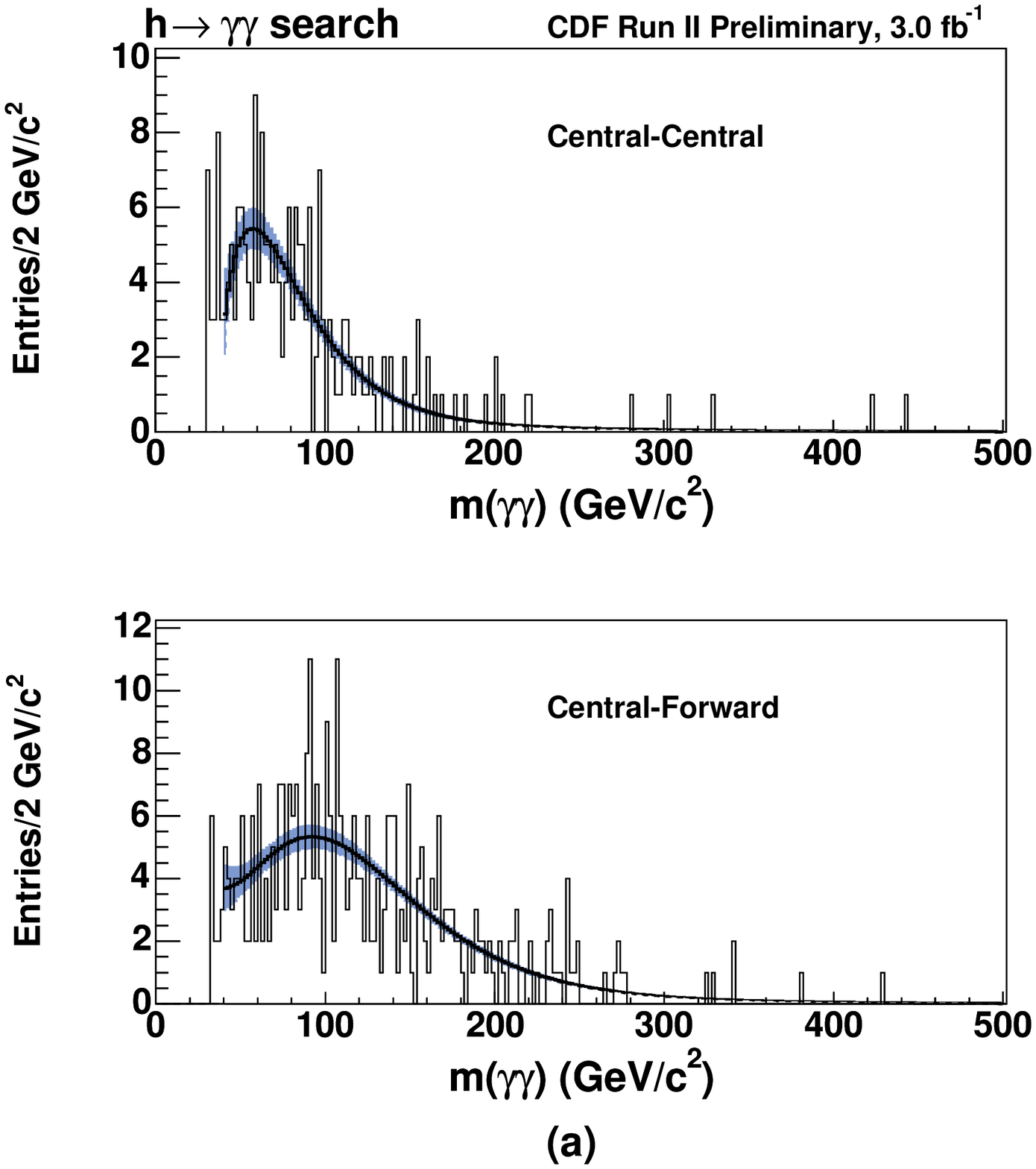}
\includegraphics[width=0.40\textwidth]{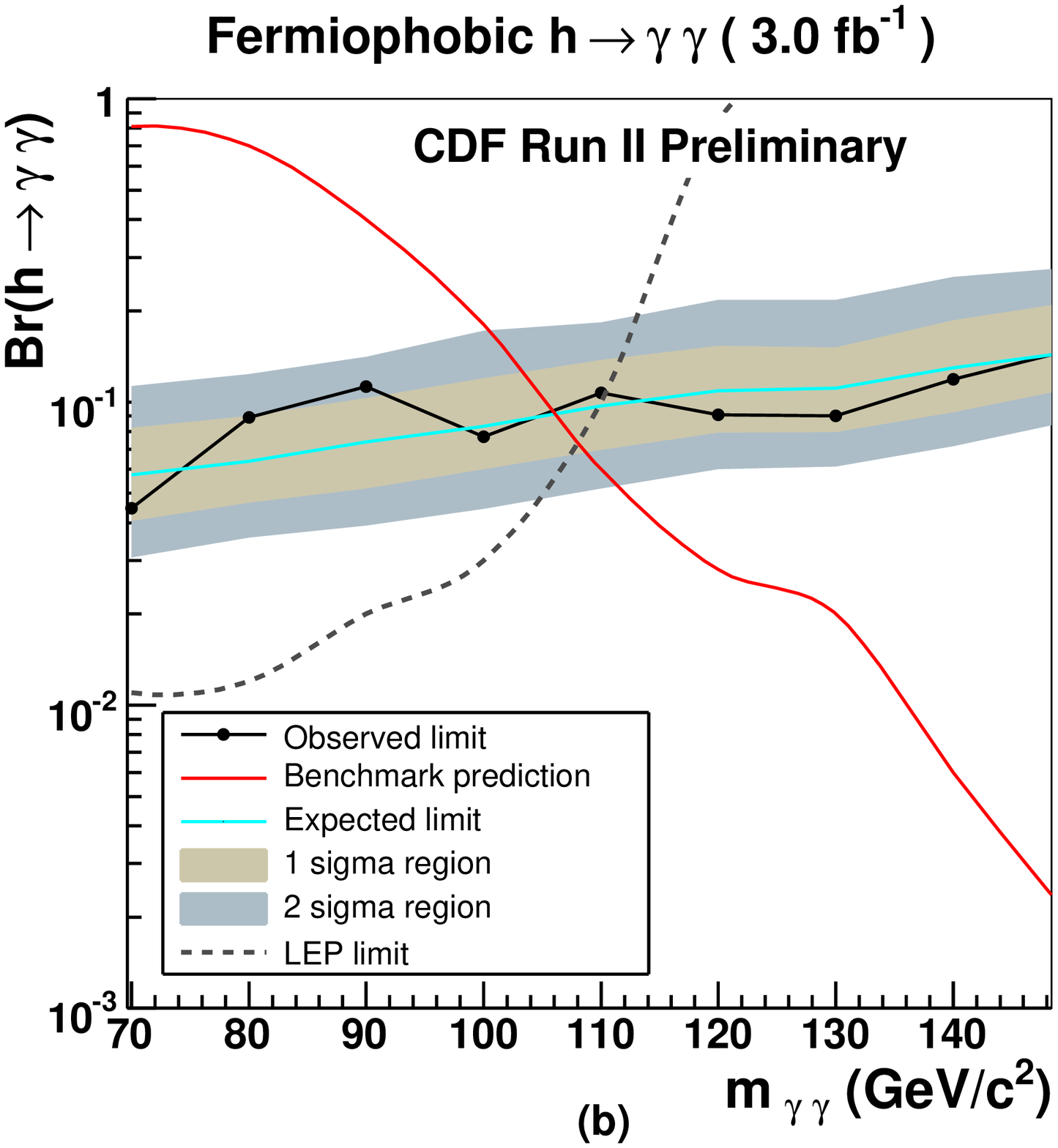}
%\subfigure[]{
%\includegraphics[width=0.40\textwidth]{figs/DataFitsErrorBands.eps}
%\label{fig:fermia}
%}
%\subfigure[]{
%\includegraphics[width=0.40\textwidth]{figs/TPeaksHiggsLimits_zoom_br_27Oct08_3.0fb.eps}
%\label{fig:fermib}
%}
\caption[]{An example of the sideband fits (a) and the 95\%
  C.L. limits (b) obtained on the \BRhgg\ from the search for a
  fermiophobic Higgs boson in the \hgg\ channel.}
\label{fig:Fermiophobic}
\end{center}
\end{figure*}

  The decay of a Higgs boson into a diphoton pair appears as a very
narrow peak in the invariant mass distribution of these two photons
($\sigma _{m}/m < 3~\%$).  The search can therfore be performed by
looking for a narrow bump on an otherwise smooth background
distribution.  No narrow resonance is observed.  In order to set
limits on the Higgs boson production rate, a sideband fit
excluding the hypothetical mass window is performed to estimate the
background in the search window (see Fig.~\ref{fig:Fermiophobic}(a)).
The analysis results in 95~\% C.L. limits on the production cross
section ($\sigma\times$\BRhgg ) and on the branching fraction (\BRhgg)
as shown in Fig.~\ref{fig:Fermiophobic}(b). The result excludes the
benchmark model predictions for \Mh\ of less than 106~\gevcc.

\section{Conclusions, and Outlook}
The CDF experiment is carefully searching for SM and BSM Higgs bosons.
A combination of results with D\O\ excludes Higgs boson masses between
160 and 170 \gevcc, and this is the first exclusion of SM
Higgs bosons based on Tevatron data.  At low mass, sensitivity is
better than three times the standard model prediction.  With more data
on tape to analyze, and improvements still being added to analysis
techniques, the Higgs boson search results will be an exciting topic until the
end of the Tevatron run.  With the full dataset expected to be on the
order of 10~\fb, there is a significant chance that the Tevatron will
see some evidence of the elusive Higgs boson.   

%------------------------------------------------------------------------------
%       Bibliography
%------------------------------------------------------------------------------
%\raggedright

\bibliographystyle{iopart-num}
\bibliography{LLWI_Higgs_group}

\end{document}